\documentclass[12pt,letterpaper]{article}

\usepackage{etoolbox}
\newtoggle{SUPPLEMENTAL}\toggletrue{SUPPLEMENTAL}
\togglefalse{SUPPLEMENTAL} 
\newcommand{\Supplemental}[2]{\iftoggle{SUPPLEMENTAL}{#1}{#2}}
\newtoggle{BLINDED}\toggletrue{BLINDED}

\Supplemental{
	\usepackage{multibib}
	\newcites{Supp}{References}
}{}

\usepackage[margin=1in,letterpaper]{geometry}
\usepackage[T1]{fontenc}
\usepackage[quotient-mode=fraction,per-mode=symbol-or-fraction]{siunitx}
\DeclareSIUnit\year{yr}
\DeclareSIUnit\mile{mi}
\DeclareSIUnit\hour{hr}
\usepackage{graphicx,grffile}
\usepackage{amsmath,amssymb,amsthm}
\usepackage{mathtools,dsfont,thmtools,mathrsfs,upgreek}
\usepackage{caption} 
\usepackage{booktabs}
\usepackage{easyReview}
\usepackage{adjustbox}
\usepackage[flushleft]{threeparttable}
\makeatletter 
\g@addto@macro\TPT@defaults{\linespread{1}\footnotesize} 
\makeatother
\usepackage{textcomp,verbatim}
\usepackage[final]{listings}
\lstdefinestyle{displayR}{language=R,frame=none,basicstyle=\ttfamily,keywordstyle=\ttfamily,stringstyle=\ttfamily,keepspaces=true,showspaces=false,showstringspaces=false,breaklines=true,upquote=true,print,columns=fullflexible}
\lstdefinestyle{inlineR}{language=R,frame=none,basicstyle=\ttfamily,keywordstyle=\ttfamily,stringstyle=\ttfamily,keepspaces=true,showspaces=false,showstringspaces=false,breaklines=true,upquote=true,print,columns=fullflexible}
\newcommand{\code}[1]{\lstinline@#1@}
\usepackage[longnamesfirst]{natbib}
\usepackage{alltt}
\usepackage[zerostyle=e,straightquotes]{newtxtt} 
\usepackage{hypernat}
\usepackage[nodisplayskipstretch]{setspace}
\usepackage{enumerate}
\usepackage{float}

\usepackage[bottom]{footmisc}
\usepackage{xr} 
\usepackage[hypertexnames=true,pagebackref=false,
bookmarks=true,pdflang=en]{hyperref}
\usepackage{bookmark}
\usepackage[capitalize,noabbrev]{cleveref}
\usepackage{multirow} 

\usepackage{tikz}
\usepackage{subcaption}
\usepackage{indentfirst}

\DeclareGraphicsExtensions{.pdf,.png}
\graphicspath{ {./Figures/} }

\Crefname{assumption}{Assumption}{Assumptions}

\crefname{equation}{}{} 
\Crefname{equation}{Equation}{Equations} 
\Crefname{method}{Method}{Methods}
\Crefname{conjecture}{Conjecture}{Conjectures}
\crefname{conjecture}{Conjecture}{Conjectures}
\Crefname{fact}{Fact}{Facts}
\crefname{fact}{Fact}{Facts}
\Crefname{enumi}{Task}{Tasks}
\crefname{enumi}{Task}{Tasks}

\theoremstyle{plain}

\theoremstyle{definition}
\newtheorem{assumption}{Assumption}
\newtheorem{definition}{Definition}

{\begin{itemize}%
	\setlength{\itemsep}{0pt}%
	\setlength{\parskip}{0pt}}%
{\end{itemize}}

{\begin{enumerate}[#1]%
	\setlength{\itemsep}{0pt}%
	\setlength{\parskip}{0pt}}%
{\end{enumerate}}

\newcommand{\R}{\mathbb{R}}

\let\originalleft\left
\let\originalright\right
\renewcommand{\left}{\mathopen{}\mathclose\bgroup\originalleft}
\renewcommand{\right}{\aftergroup\egroup\originalright}

\newcommand{\mockalph}[1]{}  

\allowdisplaybreaks[3]

\hypersetup{
colorlinks=true,
linkcolor=magenta,
citecolor=blue,
urlcolor=blue,
pdfmenubar=true,
pdftoolbar=false,
pdfstartview={FitH},
pdfauthor = {XXX},
pdfkeywords = {XXX},
pdftitle = {XXX},
pdfsubject = {XXX},
pdfpagemode={UseOutlines}
}

\begin{document}

\title{Extrapolating Away from the Cutoff in \\ Regression Discontinuity Designs}

\author{Yiwei Sun \thanks{Department of Economics, Cornell University. Email: \href{mailto:ys556@cornell.edu}{\texttt{ys556@cornell.edu}}}}

\date{November, 2023}

\maketitle

\singlespacing

\begin{abstract}
	Canonical RD designs yield credible local estimates of the treatment effect at the cutoff under mild continuity assumptions, but they fail to identify treatment effects away from the cutoff without additional assumptions. The fundamental challenge of identifying treatment effects away from the cutoff is that the counterfactual outcome under the alternative treatment status is never observed. This paper aims to provide a methodological blueprint to identify treatment effects away from the cutoff in various empirical settings by offering a non-exhaustive list of assumptions on the counterfactual outcome. Instead of assuming the exact evolution of the counterfactual outcome, this paper bounds its variation using the data and sensitivity parameters. The proposed assumptions are weaker than those introduced previously in the literature, resulting in partially identified treatment effects that are less susceptible to assumption violations. This approach accommodates both single cutoff and multi-cutoff designs. The specific choice of the extrapolation assumption depends on the institutional background of each empirical application. Additionally, researchers are recommended to conduct sensitivity analysis on the chosen parameter and assess resulting shifts in conclusions. The paper compares the proposed identification results with results using previous methods via an empirical application and simulated data. It demonstrates that set identification yields a more credible conclusion about the sign of the treatment effect.

	
	\vspace{\baselineskip}
\end{abstract}

\maketitle


\section{Introduction} \label{sec:intro}

Regression discontinuity (RD) designs have been one of the most popular empirical strategies for causal inference and program evaluation using non-experimental data since they were first introduced in \cite{thistlethwaite_campbell_1960}. In canonical sharp RD designs, each unit takes a value of the running variable. The treatment assignment rule is contingent upon whether the running variable value of a unit surpasses the specified cutoff. A unit receives treatment if its running variable exceeds the threshold; otherwise, the unit remains in the control group. This type of treatment assignment rule generates a discontinuity of the probability of receiving treatment near the cutoff, which researchers leverage to identify treatment effects. 

In the continuity framework introduced by \cite{hahn_etal_2001}, if the potential outcome functions are assumed to be continuous in the running variable at the cutoff, and the density of the running variable near the cutoff is positive, the causal effect of the treatment at the cutoff can be identified and estimated by the difference between the observed outcome functions. 
RD designs establish strong internal validity of the treatment effect for units with running variable values around the cutoff because, within a local neighborhood around the cutoff, the treatment can be considered ``as good as random'' \citep{lee_2008}. However, canonical RD estimates have limited external validity. Without additional assumptions, RD designs say little about the treatment effect for units with the running variable values away from the cutoff. Extrapolating treatment effects away from the cutoff is empirically relevant if researchers seek to understand the heterogeneous treatment effects for units with different running variable values \citep{battistin_rettore_2008, angrist_rokkanen_2015, cattaneo_keele_titiunik_2021}. Furthermore, in most empirical applications, units near the cutoff do not solely represent the policy-relevant group. Canonical RD treatment effects cannot comprehensively capture the overall effect of a policy. Extrapolating treatment effects away from the cutoff enables researchers to better assess policy implications, such as evaluating the optimality of the status quo policy and gauging welfare adjustments due to changes in the cutoff  \citep{yata_2021, zhang2022safe}. 

The fundamental challenge of identifying treatment effects away from the cutoff is that the counterfactual outcome under the alternative treatment status is never observed. An alternative interpretation of RD extrapolation is that it is essentially a missing data problem in the same spirit as the one studied in \cite{manski_1989}. The key question for the researcher is how to impute the unobserved counterfactual outcome. The goal of this paper is to provide a methodological blueprint to identify treatment effects away from the cutoff in various empirical settings by offering a non-exhaustive list of assumptions on the counterfactual outcome in both single cutoff and multi-cutoff RD designs. These assumptions aim to bound the variation of the unobserved counterfactual outcome using the data sensitivity parameters to achieve set-identified treatment effects away from the cutoff. In contrast to previous literature where strong assumptions were made to get point identification, assumptions discussed in this paper are arguably weaker and the conclusions are more robust to slight violations of the point identification assumptions. It is up to the researcher to specify the appropriate assumption to use, and it should depend on the economic context of each empirical application. It is recommended for the researcher to conduct sensitivity analysis on the choice of the bounding parameter to investigate how causal conclusions change under varying levels of tightness of the assumption. In the simulations, this paper shows that the set-identified treatment effects achieved by imposing reasonable restrictions on the variation of the counterfactual outcome can be informative about the sign of the treatment effect.

The remainder of the paper is structured as follows. Section \ref{sec:litrev} discusses the related literature; section \ref{sec:setup} introduces the setup of the problem; section \ref{sec:id} establishes the identification results under both single and multiple cutoff designs; section \ref{sec:est} demonstrates how to estimate the identification region; section \ref{sec:appsim} presents empirical and simulation results using the proposed approach; lastly, section \ref{sec:conclusion} provides concluding remarks and discusses potential future directions.


\section{Related Literature}  \label{sec:litrev}

This paper is related to the literature on causal inference, program evaluation, partial identification, and sensitivity analysis. Specifically, it contributes to the methodological literature on nonparametric identification of RD designs. Comprehensive surveys of RD designs can be found in \cite{cattaneo_titiunik_2022, choi_lee_2021, choi_lee_2017, imbens_lemieux_2008, vanderklaauw2008}. This paper aims to bring ideas in partial identification and sensitivity analysis literature to study the identification of treatment effects away from the cutoff in RD designs. 

Without any additional assumptions, canonical RD designs cannot identify treatment effects away from the cutoff. As summarized in \cite{cattaneo_keele_titiunik_2021}, there are generally two approaches to extrapolate away from the cutoff. One approach is to leverage the availability of dependent variable predictors other than the running variable, such as covariates and pre-treatment outcomes. For example, \cite{angrist_rokkanen_2015} identify treatment effects away from the cutoff in single cutoff designs by imposing a conditional independence assumption (CIA) in the spirit of \cite{imbens_rubin_2015}. Specifically, potential outcomes are assumed to be mean-independent of the running variable conditional on a set of observable covariates. \cite{palomba_2023} generalizes the result of \cite{angrist_rokkanen_2015} to multiple cutoffs, and provides a data-driven algorithm to search for a vector of covariates that satisfies CIA. Similarly, \cite{battistin_rettore_2008} and \cite{mealli_rampichini_2012} investigate extrapolation using matching-type estimators under CIA, with a specific focus on fuzzy RD designs. CIA has testable implications when imposed in RD designs \citep{angrist_rokkanen_2015}. If the properly reweighted matching-type estimates, using the distribution of covariates at the cutoff, and the local RD estimates are similar, it is suggestive that the CIA is valid \citep{angrist_rokkanen_2015, battistin_rettore_2008}. \cite{rokkanen_2015}, on the other hand, models the source of omitted variable bias in RD designs using latent factors. Assuming the running variable is one of a number of available noisy measures of the latent factors, causal effects for all values of the running variable can be nonparametrically identified. Furthermore, \cite{bennett_2020} exploits predictive covariates to explain the correlation between the running variable and the outcome of interest to extrapolate away from the cutoff.

Another general approach to extrapolate away from the cutoff is to utilize information within the research design. \cite{bertanha_imbens_2020} leverage imperfect treatment compliance in fuzzy RD designs and impose that the potential outcomes are independent of compliance types to extrapolate away from the cutoff. \cite{dong_lewbel_2015} establish nonparametric identification of the treatment effect derivative (TED), referring to the derivative of the treatment effect with respect to the running variable at the cutoff. Given the estimated TED, they can extrapolate the treatment effect at a running variable value marginally above the cutoff. In the context of multi-cutoff RD designs, \cite{cattaneo_keele_titiunik_2021} adopt a ``parallel-trend-like'' assumption on the unobserved potential outcome functions for subpopulations exposed to different cutoffs for extrapolation. All of the approaches mentioned above impose strong enough assumptions to achieve point identification of the treatment effects away from the cutoff. However, these point identification results are sensitive to slight violations of the identification assumption. Building on these previous ideas, this paper relaxes those assumptions and imposes weaker ones to yield partial identification of the treatment effect of interest. 

The partial identification approach, brought to the attention of economists firstly by works of \cite{manski_1989, manski_1990}, aims to learn about a parameter of interest given the available data and a set of assumptions. Some assumptions are more credible if they are rooted in economic theory or based on the context of the empirical application, such as demand being downward sloping \citep{manski_1997}. However, some assumptions are less credible if they are imposed only to complete a model or to achieve a desired level of precision of the problem, such as functional form and distributional assumptions \citep{molinari_2020, tamer_2010}. The philosophy of doing partial identification is to be transparent about the identifying power of different assumptions and how they shape the conclusions \citep{molinari_2020, tamer_2010}. The assumptions to identify treatment effects away from the cutoff proposed in this paper follow a top-down approach discussed in \cite{tamer_2010}, and they can be mainly summarized into two types. The first type is the bounded variation assumption directly on the counterfactual outcome. Instead of imposing parametric assumptions, this paper proposes to bound the variation of the counterfactual outcome using assumptions such as Lipchitz continuity and smoothness restrictions \citep{kim_etal_2018, kolesar_rothe_2018, yata_2021}. Additionally, in multi-cutoff designs, this paper relaxes the ``parallel-trend-like'' assumption in \cite{cattaneo_keele_titiunik_2021} by assuming the violations of parallel trends are bounded \citep{manski_pepper_2018,rambachan_roth_2023}. The other type of assumption aims to relax the conditional independence assumption that leverages pre-intervention covariates by bounding the distance between the unobserved and the observed distribution of the outcome \citep{kline_santos_2013, masten_poirier_2018}. In canonical RD designs, the role of additional covariates is mainly for estimation efficiency and falsification testing \citep{cattaneo_keele_titiunik_2023covariate, calonico_etal_2019}. However, this paper explores how the availability of additional covariates helps with the identification of treatment effects away from the cutoff. All the assumptions in this paper involve sensitivity parameters, which allow the researcher to start with tight assumptions but gradually relax them. Researchers are recommended to conduct sensitivity analysis on the chosen parameter to assess how the resulting conclusions change. In the simulations, this paper imposes a sequence of weaker assumptions and discusses when non-zero treatment effect breaks down.


\section{Setup} \label{sec:setup}

The setup of the extrapolation problem in this paper follows the conventional regression discontinuity setup in the literature \citep{cattaneo_Idrobo_Titiunik_2023, cattaneo_titiunik_2022, Lee_Lemieux_2010, imbens_lemieux_2008, vanderklaauw2008}. Assume researchers observe an independently and identically distributed sample $(Y_i, X_i, W_i, C_i, D_i), i = 1, ..., n$, where $Y_i$ is the outcome of interest, $X_i$ is the running variable (score or index), $W_i$ is a vector of observed covariates, $C_i$ is the cutoff (threshold), and $D_i$ is the binary treatment status indicator. Denote the space of covariates as $\mathcal{W}$. Assume the running variable $X_i$ has a continuous positive density $f_X(x)$ on the support $\mathcal{X}$. 

The cutoff indicator $C_i$ can be unit-specific, and takes values from a set $\mathcal{C} \subset \mathcal{X}.$ When $\mathcal{C}$ is a singleton, it is the setup of a single cutoff design, where every unit faces the same cutoff, denoted as $c_0$. When $|\mathcal{C}| \ge 2$, different units may be subject to different cutoffs, defining a multi-cutoff design. In particular, when thinking about multi-cutoff designs, this paper focuses on non-cumulative multiple cutoffs, where a unit with running variable $X_i$ can be potentially exposed to any but only one of the cutoffs $c \in \mathcal{C}$ \citep{cattaneo_Idrobo_Titiunik_2023, cattaneo_etal_2016}. Intuitively, the whole population is divided into different subpopulations based on the cutoff each subpopulation is subject to, while the support of the running variable remains common across all subpopulations. In multi-cutoff designs with non-cumulative cutoffs, it is common to assume all treated units receive the same treatment regardless of their cutoffs \citep{cattaneo_Idrobo_Titiunik_2023, cattaneo_etal_2016}. For simple exposition, assume there is only one low cutoff and one high cutoff, denoted as $\{\ell,h\}, \ell <h $.

To further simplify the problem, this paper focuses on sharp regression discontinuity designs. Consequently, the treatment assignment rule is expressed as follows:
$$D_i = \mathds{1}\{X_i \ge C_i\},$$ 
which means that a unit receives treatment if and only if its running variable is above the cutoff the unit faces. Furthermore, following the potential outcome framework in \cite{imbens_rubin_2015}, denote $Y_i(d), d \in \{ 0,1 \}$ as the potential outcomes for unit $i$. Then, the observed outcome is $Y_i = D_iY_i(1) + (1-D_i)Y_i(0).$ Lastly, denote the potential outcome regression functions for units facing cutoff $c$ with running variable value $x$ as 
$$\mu_{d,c}(x) = \mathds{E}[Y_i(d) \mid X_i = x, C_i = c], \text{ for } d \in \{ 0,1 \}, c \in \mathcal{C}.$$
Additionally, define the observed outcome regression function as 
$$\mu_{c}(x) = \mathds{E}[Y_i \mid X_i = x, C_i = c], \text{ for }c \in \mathcal{C}.$$
The potential outcome functions conditional on covariates $W_i = w$ can be defined analogously,
\begin{equation*}
	\mu_{d,c}(x, w) = \mathds{E}[Y_i(d) \mid X_i = x, C_i = c, W_i = w], \text{ for }d = \{0,1\}, c \in \mathcal{C}.
\end{equation*}

The general parameter of interest is the treatment effect of the population (or subpopulation) exposed to cutoff $c$ when the running variable takes the value $x$, which is defined as a function $\tau_c(x): \mathcal{C} \times \mathcal{X} \to \mathds{R}$, such that 
$$\tau_c(x) = \mathds{E}[Y_i(1) - Y_i(0) \mid X_i = x, C_i = c] = \mu_{1,c}(x) -\mu_{0,c}(x).$$
Assuming continuity of $\mu_{d,c}(x), d \in \{0,1\}$ within a local neighborhood of the cutoff $c$, the canonical sharp RD design can nonparametrically identify the treatment effect at the cutoff \citep{hahn_etal_2001}. For example, the treatment effect at $c_0$ in the single cutoff design is
$$\tau_{c_0}(c_0) = \mathds{E}[Y_i(1) - Y_i(0) \mid X_i = c_0, C_i = c_0] = \lim_{x \to c_0^+}\mu_{c_0}(c_0) -  \lim_{x \to c_0^-}\mu_{c_0}(c_0),$$
and in the multi-cutoff design, the treatment effect at the low cutoff $\ell$ and at the high cutoff $h$ can be separately identified as
$$\tau_{\ell}(\ell) = \mathds{E}[Y_i(1) - Y_i(0) \mid X_i = \ell, C_i = \ell] = \lim_{x \to \ell^+}\mu_{\ell}(\ell) -  \lim_{x \to \ell^-}\mu_{\ell}(\ell),$$
$$\tau_h(h) = \mathds{E}[Y_i(1) - Y_i(0) \mid X_i = h, C_i = h] = \lim_{x \to h^+}\mu_{h}(h) -  \lim_{x \to h^-}\mu_{h}(h).$$

When studying extrapolation of RD designs away from the cutoff, the specific target parameter this paper considers is the treatment effect of the treated units that are subject to cutoff $c$ but with running variable values $x$ such that $x > c$.\footnote{Similar approaches can be applied to extrapolate the treatment effect of the untreated units with running variable value below the cutoff by adapting the assumptions to bound the variation of the unobserved treated potential outcome regression functions.} Denote the specific target parameter as 
$$\theta = g(\tau_c(x)),$$ 
where $g: \mathcal{F} \to \mathds{R}$ is a known function of $\tau_c(x).$ $\mathcal{F}$ is the functional space that $\tau_c(x)$ belongs to, which later will be specified depending on the restrictions imposed on the potential outcome functions. If the researcher is interested in the treatment effect of units at a particular value of the running variable $x^*$, $g(\cdot)$ is the function such that 
$$\theta = \tau_c(x^*),  \text{ for a particular } x^*.$$
Alternatively, if the researcher is interested in the average treatment effect of treated units with running variable in the range $[a,b], a, b \in \mathcal{X}$ above the cutoff \citep{angrist_rokkanen_2015}, then $g(\cdot)$ is the integral function such that 
$$\theta = \int_a^b \tau_c(x) f_{X| x \in [a,b]}(x)dx,$$
where $f_{X| x \in [a,b]}(x)$ is the density of $x$ conditional on $x\in [a,b]$.


\section{Identification} \label{sec:id}

This paper focuses on discussing possible assumptions needed to extrapolate to the right of the cutoff, specifically the treatment effects on the treated units with running variable values higher than the cutoff. In single cutoff designs, the target parameter is $g(\tau_{c_0}(x)), x \ge c_0$; and in multi-cutoff designs, the target parameter is the treatment effect of the subpopulation subject to the low cutoff, but with running variable value between $\ell$ and $h$, $g(\tau_{\ell}(x)), x \in (\ell, h)$. The fundamental identification challenge is that the untreated potential outcome functions are never observed. The general approach is to impose restrictions to bound the variation of the untreated potential outcome functions above the cutoff. Section \ref{subsec:single} discusses assumptions in single cutoff designs that can be directly imposed on the evolution of the untreated potential outcome functions. Assumptions in single cutoff designs can also be used for extrapolation in multi-cutoff designs; furthermore, section \ref{subsec:multi} discusses additional assumptions that can be imposed in multi-cutoff designs that exploit the existence of other subpopulations subject to different cutoffs. 


\subsection{Single Cutoff Designs} \label{subsec:single}

To restrict the variation of the untreated potential outcome function, this paper proposes to specify $\mu_0(x) \in \mathcal{U}(\kappa)$, where $\mathcal{U}(\kappa)$ satisfy some properties that the researcher deems to be desirable depending on the empirical application, and $\kappa$ is a sensitivity parameter. By construction, the identification region of the treatment effect away from the cutoff is all the possible differences between the observed treated outcome and the restricted untreated potential outcome. Formally, 
\begin{equation}
	\begin{split}
		\Uptheta(\mathcal{U}) := \Big\{ \theta \in \mathds{R} : \;\; &\exists \mu_{0, c_0}(x) \in \mathcal{U}, \tau_{c_0}: \mathcal{X} \to \mathds{R}
		\text{ s.t. } \\
		& \theta = g(\tau_{c_0}(x)), \tau_{c_0}(x)  = \mu_{c_0}(x) - \mu_{0,c_0}(x), \text{ 
			for  } x \ge c_0 \Big\}.
	\end{split}  
	\label{single id}
\end{equation}
Next, I discuss some possible choices of $\mathcal{U}$.

\subsubsection{Lipschitz Continuity} \label{subsubsec:lipcont}
One type of restriction is to assume the untreated potential outcome function is Lipschitz continuous with a Lipschitz constant $\kappa$. The Lipschitz continuity constraint bounds the maximum possible change in $\mu_{0,c_0}(x)$ given a shift in the running variable $x$ by one unit \citep{yata_2021}. If the researcher is willing to additionally assume $\mu_{0,c_0}(x)$ is differentiable, the Lipschitz restriction is imposing the absolute value of the derivative of $\mu_{0,c_0}(x)$ is bounded by $\kappa$. Formally, this restriction is stated as 
\begin{equation}
	\mathcal{U}^{Lip}(\kappa) =\Big \{\mu_{0, c_0}(x): |\mu_{0, c_0}(x) - \mu_{0, c_0}(\Tilde{x})| \le \kappa |x - \Tilde{x}|, \forall x, \Tilde{x} \in \mathcal{X} \Big\}.
\end{equation}
If the researcher has access to additional pre-intervention covariates, the Lipschitz constraint can also extend to potential outcome functions conditional on these covariates such that 
\begin{equation}
	\begin{split}
		\mathcal{U}_w^{Lip}(\kappa) = \Big\{\mu_{0, c_0}(x, w): |\mu_{0, c_0}(x, w) - \mu_{0, c_0}(\Tilde{x}, \Tilde{w})| &\le \kappa ||(x, w) - (\Tilde{x}, \Tilde{w})||, \\
		&\forall (x, w), (\Tilde{x}, \Tilde{w}) \in \mathcal{X} \times \mathcal{W} \Big\}.
	\end{split}
\end{equation}
When the covariates $W_i$ are finite and discrete, the canonical sharp RD treatment effect $\tau_{c_0}(x^*)$ at running variable value $x^*$ becomes a weighted average of the conditional RD treatment effects of each subgroup defined by $W_i  = w$, with weights related to the conditional distribution of $W_i \mid X_i$ at $x^*$ \citep{cattaneo_keele_titiunik_2023covariate}. The RD treatment effects conditional on each subgroup can be recovered by running a fully saturated polynomial regression with subgroup indicators, and can potentially be asymptotically more efficient under the correct implementation \citep{calonico_etal_2019}. When the covariates are continuous and high-dimensional, it may be infeasible to estimate the average treatment effect for each subgroup; instead, a solution is to estimate conditional ATE for a large number of covariate partitions without specifying subgroups a priori using machine learning methods \citep{cattaneo_keele_titiunik_2023covariate}. 

\subsubsection{Smoothness Restriction} \label{subsubsec:singleSR}
Similarly, in addition to restricting the rate of change, the researcher can impose smoothness restriction on the untreated potential outcome function \citep{kolesar_rothe_2018}. Assume $\mu_{0, c_0}(x)$ is twice differentiable, define the class of untreated potential outcome functions with bounded second derivatives, $\mathcal{U}^{SD}(\kappa)$, as 
\begin{equation}
	\mathcal{U}^{SD}(\kappa) = \Big\{\mu_{0, c_0}(x): |\mu_{0, c_0}'(x) - \mu_{0, c_0}'(\Tilde{x})| \le \kappa |x - \Tilde{x}|,    \forall x, \Tilde{x} \in \mathcal{X} \Big\}.
\end{equation}
Analogously, given the availability of pre-intervention covariates, the researcher can impose the smoothness restriction conditional on covariates $W_i = w.$ The analysis goes through in the same spirit as in section \ref{subsubsec:lipcont}. This type of restriction can be extended to bound higher-order variation of the potential outcome functions with strengthened assumptions on the differentiability of $\mu_{0, c_0}(x).$

\subsubsection{Bounded Deviation from CIA} \label{subsubsec:CIA}

Instead of directly restricting the variation of the untreated potential outcome function, another type of popular assumption to extrapolate away from the cutoff is the "ignorability" conditional on observables assumption, which assumes that conditional on observable pre-intervention covariates, the potential outcome is independent or mean independent of the running variable \citep{angrist_rokkanen_2015, mealli_rampichini_2012, battistin_rettore_2008}. Another interpretation of the conditional independence assumption is that the untreated potential outcome above the cutoff is missing at random. The "ignorability" conditional on observables is untestable, making its resulting causal conclusion susceptible to even small violations of this assumption. Instead of imposing that conditional independence holds exactly, this paper proposes to bound the deviation from conditional independence, or equivalently, the nonignorable selection on observables. 

Following similar ideas in \cite{kline_santos_2013} and \cite{masten_poirier_2018}, this paper bounds the deviation from conditional independence using the maximal Kolmogorov-Smirnov distance. Firstly, notice that the untreated potential outcome function is observed for $x \le  c_0$. Next, define the probability of $\mu_{0,c_0}(x,w)$ being observed as 
\begin{equation*}
	P_{c_0}(w) = P(x \le c_0 \mid W = w).
\end{equation*}
The distribution of interest is the distribution of $Y(0)$ conditional on covariate $W= w$ , denoted as $F_{Y(0) \mid w}(y)  = P(Y(0) \le y \mid W = w)$. The researcher observes this distribution for $x \le c_0$, denoted as $F_{Y(0) \mid 0,w}(y) = P(Y \le y \mid  X \le c_0, W = w)= P(Y \le y \mid D = 0, W = w)$. However, the researcher does not observe the distribution of $Y(0)\mid w$ for $x>c_0$, denoted as $F_{Y(0) \mid 1,w}(y) = P(Y \le y \mid  X > c_0, W = w)= P(Y \le y \mid D = 1, W = w)$. The Kolmogorov-Smirnov (KS) distance is defined as 
\begin{equation}
	\sup_{w \in \mathcal{W}} \sup_{y \in \R} |F_{Y(0) \mid 0,w}(y) - F_{Y(0) \mid 1,w}(y) |. 
\end{equation}
The conditional independence assumption is stated as $Y(0) \bot X \mid W$, which implies $F_{Y(0) \mid 0,w}(y) = F_{Y(0) \mid 1, w}(y)$; or equivalently, the KS distance between $F_{Y(0) \mid 0,w}(y) $ and $F_{Y(0) \mid 1,w}(y)$ is zero. Instead of assuming the KS distance is exactly zero, this paper assumes that the distribution of $Y(0) \mid 1,w$ belongs to the class of distributions with no more than $\kappa$ KS distance away from the observed distribution $Y(0) \mid 0,w$, denoted as 
\begin{equation}
	\mathcal{Y}(\kappa) =\Big\{ F_{Y(0) \mid 1,w}(y): \sup_{w \in \mathcal{W}} \sup_{y \in \R} |F_{Y(0) \mid 0,w}(y) - F_{Y(0) \mid 1,w}(y) | \le \kappa \Big\},
\end{equation}
where $\kappa$ is a sensitivity parameter. 

\cite{kline_santos_2013} discuss one intuitive interpretation of the sensitivity parameter $\kappa$ in this case. To put in perspective, consider $F_{Y(0) \mid 1,w}(y)$ as a mixture model such that 
\begin{equation*}
	F_{Y(0) \mid 1,w}(y) = (1-\kappa) F_{Y(0) \mid 0,w}(y) + \kappa \Tilde{F}_{Y(0) \mid w} (y),  \forall w \in \mathcal{W}, y \in \R,
\end{equation*}
where $\Tilde{F}_{Y(0) \mid w}$ is any unknown distribution. Then, the KS distance can be re-written as 
\begin{equation*}
	\begin{split}
		&\sup_{w \in \mathcal{W}} \sup_{y \in \R} |F_{Y(0) \mid 0,w}(y) - (1-\kappa) F_{Y(0) \mid 0,w}(y) - \kappa \Tilde{F}_{Y(0) \mid w}(y) | \\
		&= \kappa \underbrace{\sup_{w \in \mathcal{W}} \sup_{y \in \R} |F_{Y(0) \mid 0,w}(y) - \Tilde{F}_{Y(0) \mid w}(y) |}_{\in [0,1]}  \in [0,\kappa].
	\end{split}
\end{equation*}
Hence, imposing $ F_{Y(0) \mid 1,w}(y) \in \mathcal{Y}(\kappa)$ is equivalent to assuming that there is at most $\kappa$ fraction of the unobserved untreated potential outcome not well represented in the observed data distribution conditional on covariates. Connecting to the foundational work of \cite{manski_2003}, bounds on $F_{Y(0) \mid w}(y) $ can be derived using the law of total probability,
\begin{equation}
	\begin{split}
		F_{Y(0) \mid w}(y) &= P_{c_0}(w)F_{Y(0) \mid 0,w}(y) + \Big(1-P_{c_0}(w)\Big)\Big( (1-\kappa) F_{Y(0) \mid 0,w}(y) + \kappa \Tilde{F}_{Y(0) \mid w}(y) \Big)\\
		&= \Big( 1+\kappa(P_{c_0}(w)-1) \Big) F_{Y(0) \mid 0,w}(y) + \Big(1-P_{c_0}(w)\Big) \kappa \underbrace{\Tilde{F}_{Y(0) \mid w}(y)}_{\in [0,1]}\\
		\Rightarrow F_{Y(0) \mid w}(y) &\in \Bigg[ \Big( 1+\kappa(P_{c_0}(w)-1) \Big) F_{Y(0) \mid 0,w}(y), \Big( 1+\kappa(P_{c_0}(w)-1) \Big) F_{Y(0) \mid 0,w}(y) + \Big(1-P_{c_0}(w)\Big) \kappa \Bigg].
	\end{split}
\end{equation}
Given bounds on $F_{Y(0) \mid w}(y)$ and the specified support of $Y(0)$, the researcher can back out bounds on the expectation of $Y(0)$ at a particular value of the running variable conditional on covariates, and therefore partially identify the treatment effect away from the cutoff.


\subsection{Multi-Cutoff Designs} \label{subsec:multi}

This section focuses on discussing the possible extrapolation assumptions in multi-cutoff RD designs. Firstly, it is worth noting that to extrapolate away from the cutoff in non-cumulative multi-cutoff designs,  all the assumptions for single cutoff designs still apply because researchers can view the extrapolation problem for each subpopulation subject to different cutoffs as a separate single cutoff extrapolation problem. Additionally, due to the existence of other subpopulations with different cutoffs, there is more information that researchers can leverage for extrapolation. 

\begin{figure}
	\centering
	\includegraphics[width=10cm]{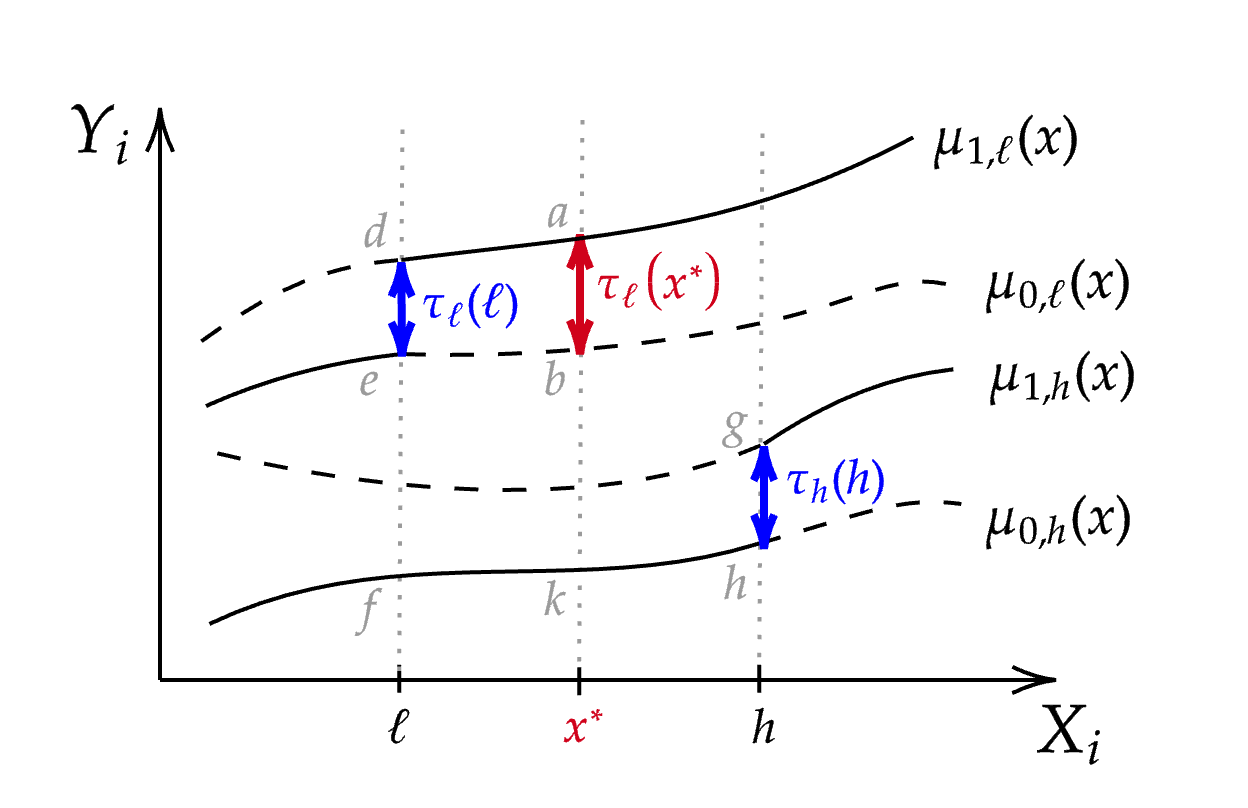}
	\caption{Multiple Cutoff Regression Discontinuity Design}
	\label{fig:multi}
\end{figure}

\cite{cattaneo_keele_titiunik_2021} use a ``parallel-trend-like'' assumption to extrapolate away from the cutoff in multi-cutoff designs. Without loss of generality, consider two cutoffs, $\{ \ell, h \}$, with the goal to extrapolate the treatment effects, $g(\tau_{\ell}(x))$, for the subpopulation subject to cutoff $\ell$ but with running variable value $x \in (\ell, h)$. Refer to Figure \ref{fig:multi} as an example. Standard multi-cutoff RD designs can nonparametrically identify $\tau_{\ell}(\ell) = \lim_{x \to \ell^+}\mu_{\ell}(\ell) -  \lim_{x \to \ell^-}\mu_{\ell}(\ell)$ and $\tau_{h}(h) = \lim_{x \to h^+}\mu_{h}(h) -  \lim_{x \to h^-}\mu_{h}(h)$, which are the $de$ and $gh$ segment in Figure \ref{fig:multi}. The identification challenge is that $\mu_{0,\ell}(x), x \in (\ell,h)$, is never observed. Without loss of generality, consider identifying the treatment effect at $x^*.$ What the researcher can observe in the data is the difference between the treated outcome function for units facing the lower cutoff and the untreated outcome function for units facing the higher cutoff at $x^*$, denoted as $\mu_{1,\ell}(x^*) - \mu_{0,h}(x^*)$, which corresponds to $ak$ in Figure \ref{fig:multi}. If the difference between the untreated potential outcomes for units facing the lower cutoff, $\mu_{0,\ell}(x^*)$, and the untreated potential outcome for units facing the higher cutoff, $\mu_{0,h}(x^*)$, can be quantified, the researcher can identify treatment effect at $x^*$, denoted as $$\tau_{\ell}(x^*) = \Big(\mu_{1,\ell}(x^*) - \mu_{0,h}(x^*)\Big) - \Big(\mu_{0,\ell}(x^*) - \mu_{0,h}(x^*)\Big) = \mu_{1,\ell}(x^*) -\mu_{0,\ell}(x^*).$$

The difference between untreated potential outcomes for units with the same running variable value but different cutoffs can be interpreted as bias driven by both observable and unobservable predictors of the outcome other than the running variable across subpopulations \citep{cattaneo_keele_titiunik_2021}. Formally, bias is defined as the following.
\begin{definition}{(Bias)}	\label{def:bias}
	\emph{
		Define bias as a function $B:\mathcal{X} \to \mathds{R}$ such that $$B(x) = \mu_{0,\ell}(x) - \mu_{0,h}(x).$$}
\end{definition}

Under assumption \ref{ass:continuity}, where the potential outcome functions are continuous not only at the cutoff but also over the whole extrapolation interval $[\ell, h]$, and assumption \ref{ass:constantbias}, which is equivalent to imposing $\mu_{0,\ell}(x)$ and $\mu_{0,h}(x)$ are parallel,  \cite{cattaneo_keele_titiunik_2021} establish point identification of $\tau_{\ell}(x^*)$ as 
\begin{equation} \label{res:ptid}
	\tau_{\ell}(x^*) = \mu_{\ell}(x^*) - [\mu_h(x^*) + B(\ell)].
\end{equation}

\begin{assumption}{(Continuity)} \label{ass:continuity}
	\emph{
		$\mu_{d,c}(x)$ is continuous in $x \in [\ell,h]$ for $d = 0, 1$ and for all $c$.}
\end{assumption}

\begin{assumption}{(Constant bias)} \label{ass:constantbias}
	\emph{  
		$B(\ell) = B(x)$, for all $x \in (\ell, h)$.
	}
\end{assumption}

Notably, this constant bias assumption is similar to the parallel trend assumption in difference-in-difference (DiD) designs, both of which are untestable assumptions. Even if the researcher observes that the untreated potential outcome functions across subpopulations are parallel for running variable values below the low cutoff, this does not guarantee that the bias remains constant above the low cutoff. If the true bias is not exactly constant, the identification result fails. This issue resembles the pre-trend testing concerns raised by \cite{roth_2022}. 

To derive more robust identification results, this paper aims to weaken the constant bias assumption. Instead of assuming the bias is exactly constant, this paper assumes that the variation of the unobserved bias can be bounded based on the bias observed in the data \citep{manski_pepper_2018, rambachan_roth_2023}. This paper assumes that the bias function $B \in \mathscr{B}$, where $\mathscr{B}$ is a collection of functions satisfying some desired restrictions. To study the identification problem, this paper highlights the following causal decomposition of the observables in the data. Firstly, define the observables as a piece-wise function $\gamma(x)$, where 
$$\gamma(x) = 
\begin{cases}
	\mu_{0,\ell}(x)-\mu_{0,h}(x) & \text{for } x \le \ell \\
	\mu_{1,\ell}(x)-\mu_{0,h}(x) & \text{for } x \in (\ell,h) \\
	\mu_{1,\ell}(x)-\mu_{1,h}(x) & \text{for } x \ge h .
\end{cases}$$
For an illustration, see Figure \ref{fig:multi}. For $x \le \ell$, both subpopulations are untreated and the researcher observes the difference between the untreated potential outcome functions. For $x \in (\ell, h)$, the subpopulation exposed to cutoff $\ell$ is treated, but the subpopulation exposed to cutoff $h$ is not. Lastly, for $x \ge h$, both subpopulations are treated. Since I only consider extrapolating treatment effects in the region $(\ell, h)$, I focus the analysis on $x <h$ onward. Then the observables can be decomposed as the sum of the causal parameter of interest and a bias term:
\begin{equation}
	\gamma(x) = 
	\begin{cases}
		0 + B(x) & \text{for } x \le \ell \\
		& \\
		\tau_{\ell}(x) + B(x) & \text{for } x \in (\ell,h),
	\end{cases}
	\label{decomposition}
\end{equation}
where $\tau_{\ell}(x) = \mu_{1,\ell}(x) - \mu_{0,\ell}(x),$ and $B(x) = \mu_{0,\ell}(x) - \mu_{0,h}(x)$, as defined before. If the running variable is below the cutoff $\ell$, no treatment is given; therefore, the causal parameter of interest is mechanically zero, and all that is observed in the data is the bias. If the running variable exceeds the cutoff, the difference between observed outcome functions is the sum of the causal effect and the bias. 

The partially identified treatment effect $\theta$ is the set of values for $\theta$ that are consistent with the observables, $\gamma(x)$, under the restriction $B\in \mathscr{B}$,
\begin{equation}
	\begin{split}
		\Uptheta(\gamma, \mathscr{B}) :=\Bigg\{ \theta \in \mathds{R}: \exists B &\in \mathscr{B}, \tau_{\ell}(x):\mathcal{X} \to \mathds{R} \\
		&\text{ s.t. } \theta = g(\tau_{\ell}(x)), \text{ where } x \in (\ell,h), \text{ and} \\
		&\;\;\;\;\;\;\;  \gamma(x) = 
		\begin{cases}
			B(x) & \text{for } x \le \ell \\
			\tau_{\ell}(x) + B(x) & \text{for } x \in (\ell,h)
		\end{cases} \Bigg\}.
	\end{split}
\end{equation}
The class of bias functions $\mathscr{B}$ needs to be specified by the researcher, and the choice of $\mathscr{B}$ should depend on the context of each empirical application and the available data, $\gamma(x), x \le \ell$. Under this framework, this paper discusses several restrictions researchers can impose based on the idea of bounded variation in \cite{manski_pepper_2018} and \cite{rambachan_roth_2023}.

\subsubsection{Bounded Absolute Magnitudes} \label{subsubsec:BAM}

In some empirical applications, researchers may be willing to assume that the magnitude of the bias across running variable values has bounded variation. Intuitively, this assumption suggests that in the data, the bias function oscillates within a band. One possible restriction is to impose the bias function $B \in \mathscr{B}^{AM}(\kappa)$ for $\kappa \ge 0$, where
\begin{equation}
	\begin{split}
		\mathscr{B}^{AM}(\kappa) = \Big\{ B \in \mathbf{C}^0: \; &\forall x \le \ell, B(x) = \gamma(x), \\
		&\forall x \in (\ell,h), |B(x)| \le \kappa \cdot \bar{B} \Big\},
	\end{split}
\end{equation}
with $\bar{B} = \max_{\Tilde{x} \le \ell}|B(\Tilde{x})| = \max_{\Tilde{x} \le \ell}|\gamma(\Tilde{x})|$ being a constant that can be estimated and $\kappa$ as a sensitivity parameter. When the researcher is willing to believe that the absolute magnitude of the bias for the running variable taking values above the threshold is no bigger than the biggest absolute magnitude bias they observe (bias for the running variable taking values below the threshold), a natural benchmark is to let $\kappa = 1.$

\subsubsection{Bounded Relative Magnitudes} \label{subsubsec:BRM}

In some applications, it may be reasonable to assume the rate of change of the bias above cutoff $\ell$ is close to the rate of change of the bias below $\ell$. Then, the researcher can impose restrictions to bound the first derivatives of $B$. Firstly, assume $B$ is continuously differentiable; equivalently, $\mu_{d,\ell}(x), d \in \{0,1\}$, is continuously differentiable. Then the class of bias functions with bounded first derivatives belongs to $\mathscr{B}^{RM}(\kappa)$ for sensitivity parameter $\kappa \ge 0$, where
\begin{equation}
	\begin{split}
		\mathscr{B}^{RM}(\kappa) = \Big\{ B \in \mathbf{C}^1: \; &\forall x \le \ell, B(x) = \gamma(x), \\
		&\forall x \in (\ell,h), |B'(x)| \le \kappa \cdot  \Bar{B'} \Big\}, 
	\end{split}
\end{equation}
and $\Bar{B'} = \max_{\Tilde{x} \le \ell}|B'(\Tilde{x})| = \max_{\Tilde{x} \le \ell}|\gamma'(\Tilde{x})|. $ In particular, if the researcher observes the bias below $\ell$ to be approximately linear, it may be reasonable to let $\kappa = 1$ and assume the deviation from the linear trend of the bias above $\ell$ is not too large. 

\subsubsection{Smoothness Restriction} \label{subsubsec:multiSR}

Depending on the context of the problem, even though the trend of the bias function is not linear, it may still be reasonable to assume that the evolution of bias is smooth after passing the lower threshold. Assume that $B$ is twice continuously differentiable; equivalently, $\mu_{d,\ell}(x), d \in \{0,1\}$ is twice continuously differentiable. The class of smooth bias functions, $\mathscr{B}^{SD}(\kappa)$, given sensitivity parameter $\kappa \ge 0$, can be defined as
\begin{equation}
	\begin{split}
		\mathscr{B}^{SD}(\kappa) = \Big\{ B\in \mathbf{C}^2: \; &\forall x \le \ell, B(x) = \gamma(x), \\
		&\forall x \in (\ell,h), |B''(x)| \le \kappa \cdot  \Bar{B}'' \Big\},
	\end{split}
\end{equation}
where $\Bar{B''} = \max_{\Tilde{x} \le \ell}|B''(\Tilde{x})| = \max_{\Tilde{x} \le \ell}|\gamma''(\Tilde{x})|. $ This restriction imposes that the unobserved second derivatives of the bias are not bigger than $\kappa$ times the maximal second derivative observed in the data.

\subsubsection{Bounded Polynomial Expansion} \label{subsubsec:BPE}

A generalization of the restrictions mentioned above is to bound higher-order variation of the bias function. Assume $B$ is $p$-times continuously differentiable; equivalently, $\mu_{d,\ell}(x), d \in \{0,1\}$ is $p$-times continuously differentiable, for some $p \in \{0,1,2, \cdots\}$. As shown in the appendix of \cite{cattaneo_keele_titiunik_2021}, the bias functions can be approximated using a polynomial expansion as 
\begin{equation*}
	B(x) = \sum_{s =0}^p \frac{1}{s!} B^{(s)}(\ell) \cdot (x-\ell)^s, \text{ for } x \in (\ell, h),
\end{equation*}
where $B^{(s)}(x) = \mu^{s}_{0,\ell}(x) -  \mu^{s}_{0,h}(x)$ and $\mu^{s}_{d,c}(x) = \frac{\partial^{s} \mu_{d,c}(x) }{\partial x^s}$. The researcher needs to specify $p$, and the choice of $p$ can depend on the optimal polynomial order approximation of the observed bias function for $x \le \ell.$ Then, the class of bias functions with bounded higher order derivatives, for $s \in \{0,1, \cdots, p\}$ and given sensitivity parameters $\kappa_s \ge 0$, can be defined as
\begin{equation}
	\begin{split}
		\mathscr{B}^{BPE}(\kappa) = \Big\{ B\in \mathbf{C}^p: \; &\forall x \le \ell, B(x) = \gamma(x), \\
		&\forall x \in (\ell,h), |B^{(s)}(x)| \le \kappa_s \cdot  \Bar{B}^{(s)},  s \in \{0,1, \cdots, p\}\Big\},
	\end{split}
\end{equation}
where $\kappa_s$ are sensitivity parameters that can vary by $s$, and $\Bar{B}^{(s)} = \max_{\Tilde{x} \le \ell}|B^{(s)}(\Tilde{x})| = \max_{\Tilde{x} \le \ell}|\gamma^{(s)}(\Tilde{x})|$ is the maximum $s$-th derivative of $B(x)$ for $x \le \ell.$ Researchers can also let $\Bar{B}^{(s)} = 1$ to directly impose the $s$-th derivative of the bias function is no bigger than a constant $\kappa_s$.

\subsubsection{Intersection Bounds} \label{subsubsec:ensemble}

Another way to incorporate all the assumptions discussed so far is to specify $\mathscr{B}^{IB}(\kappa)$ to be any intersection of the classes of bias functions mentioned above. The bounds of the identification region achieved by imposing $B \in \mathscr{B}^{IB}(\kappa)$ will be the intersection bounds of the identification region achieved by imposing $B$ belongs to each of the class of functions that defines $\mathscr{B}^{IB}(\kappa)$. For example, the researcher may be willing to impose that both the absolute and relative magnitudes of the bias function are bounded in the sense discussed in section \ref{subsubsec:BAM} and \ref{subsubsec:BRM}. Assuming that $B$ is continuously differentiable in this case, the class of bias functions belongs to $\mathscr{B}^{IB}(\kappa) \equiv \mathscr{B}^{AM}(\kappa) \cap \mathscr{B}^{RM}(\kappa) $, for sensitivity parameter $\kappa \ge 0$, where
\begin{equation}
	\begin{split}
		\mathscr{B}^{IB}(\kappa)  = \Big\{ B\in \mathbf{C}^1: \; &\forall x \le \ell, B(x) = \gamma(x), \\
		&\forall x \in (\ell,h), |B(x)| \le \kappa \cdot \bar{B}, \\
		& \;\;\;\;\;\;\;\;\;\;\;\text{and  } |B'(x)| \le \kappa \cdot  \Bar{B'} \Big\},
	\end{split}
\end{equation}
with $\bar{B}$ and $\Bar{B'}$ defined similarly as before.

\section{Estimation} \label{sec:est}

The identification region given the restrictions discussed above is easy to estimate using nonparametric methods, such as series regression and kernel density estimation. Under some restrictions, it is also possible to derive a closed-form characterization of the identification region. As an example, I discuss estimating the identification region in multi-cutoff designs under the bounded absolute and relative magnitudes assumption in this section, which will be implemented in section \ref{sec:appsim} to showcase the empirical illustration. 

To start off, let the parameter of interest be the treatment effect at a particular running variable value $x^*$ above the low cutoff, denoted as $\tau_{\ell}(x^*)$. Imposing $B \in \mathscr{B}^{AM}(\kappa)$ is essentially putting restrictions on $\mu_{0,\ell}(x)$ such that
\begin{equation*}
	\begin{split}
		&|B(x)| \le \kappa \bar{B}, \forall x \in (\ell,h)\\
		\Leftrightarrow & 	-\kappa \bar{B} \le \mu_{0,\ell}(x) - \mu_{0,h}(x) \le \kappa \bar{B}, \forall x \in (\ell,h)\\
		\Leftrightarrow & \mu_{0,h}(x)-\kappa \bar{B} \le \mu_{0,\ell}(x) \le \mu_{0,h}(x)+\kappa \bar{B}, \forall x \in (\ell,h),
	\end{split}
\end{equation*} 
where $\kappa$ is the sensitivity parameter specified by the researcher, and $\mu_{0,h}(x)$ and $\bar{B}$ can both be estimated from the data using series regression. For the empirical illustration in section \ref{sec:appsim}, this paper uses the power basis $x^{j-1}, j \in \{1, 2, \cdots, J\}$.  The optimal order of the polynomials, $J^*$, is chosen to be the one that minimizes the leave-one-out cross-validation criterion, which is shown to be a nearly unbiased estimator of mean-squared-forecast-error (MSFE) \citep{hansen_2014}. Hence, the estimated bounds on $\mu_{0,\ell}(x)$ at $x^*$ are simply the ones with the estimated $\hat{\mu}_{h}(x^*)$ and $\hat{\bar{B}}$,
\begin{equation*}
	\mu_{0,\ell}(x^*) \in \Big[ \hat{\mu}_{h}(x^*) - \kappa \hat{\bar{B}},  \hat{\mu}_{h}(x^*)+ \kappa \hat{\bar{B}}\Big].
\end{equation*}
Given the bounds on $\mu_{0,\ell}(x)$, and the construction of the identification region, the estimated treatment effect $\tau_{\ell}(x^*)$ is 
\begin{equation}
	\Uptheta(\hat{\gamma}, \mathscr{B}^{AM}(\kappa)) = \Bigg[ \hat{\mu}_{\ell}(x^*) - \Big(\hat{\mu}_{h}(x^*)+ \kappa \hat{\bar{B}}\Big), \hat{\mu}_{\ell}(x^*) - \Big(\hat{\mu}_{h}(x^*) - \kappa \hat{\bar{B}}\Big) \Bigg],
\end{equation}
where $\hat{\mu}_{\ell}(x^*)$ is estimated in the same way using series regression with polynomial basis functions.

Similarly, by imposing $B \in \mathscr{B}^{RM}(\kappa)$, $B(x^*)$ attains its maximum above $\ell$ when $B(x)$ strictly increases at the rate of $\kappa \bar{B}'$, and attains its minimum above $\ell$ when $B(x)$ strictly decreases at the rate of $\kappa \bar{B}'$. Hence, the estimated bounds on the bias are the following 
\begin{equation*}
	B(x^*) \in \Big[ \hat{B}(\ell) - \kappa\hat{\Bar{B}}'(x^* - \ell),\hat{B}(\ell) + \kappa\hat{\Bar{B}}'(x^* - \ell)\Big],
\end{equation*}
where all quantities are either estimated or given by the researcher. Lastly, by construction, the identification region is 
\begin{equation}
	\Uptheta(\hat{\gamma}, \mathscr{B}^{RM}(\kappa)) = \bigg[\hat{\gamma}(x^*) - \Big(\hat{B}(\ell) + \kappa\hat{\Bar{B}}'(x^* - \ell)\Big), \hat{\gamma}(x^*) - \Big( \hat{B}(\ell) - \kappa\hat{\Bar{B}}' (x^* - \ell) \Big)\bigg].
\end{equation}
Following the same logic, closed-form bounds of the identification region under high-order restrictions can be attained by backward induction. 


\section{Empirical Illustration and Simulations} \label{sec:appsim}

In this section, I present two examples of the partially identified treatment effect away from the cutoff in multi-cutoff RD designs. The first example uses the empirical application in \cite{cattaneo_keele_titiunik_2021}. The second example uses the simulated data set 1 in the \texttt{rdmulti} package.\footnote{The simulated data set can be accessed at \href{https://rdpackages.github.io/rdmulti/}{rdpackages.github.io/rdmulti}.} In both examples, I extrapolate the treatment effect to one particular running variable value above the lower cutoff, and I compare the identification results when imposing the bounded absolute magnitudes and the bounded relative magnitudes restriction with varying values of the sensitivity parameter, discussed in section \ref{subsubsec:BAM} and \ref{subsubsec:BRM}. 

\subsection{ACCES subsidized loan program}

ACCES is a subsidized loan program administered by the Colombia Institute for Educational Loans and Studies Abroad (ICETEX) to provide tuition credits to underprivileged populations. The eligibility criterion for this program depends on whether an individual's exam position score on a high school exit test, known as SABER 11, passes the specified cutoff. The position score of a student is their ranking out of 1000 quantiles of the SABER 11 scores among all students who took the exam that semester.  To be consistent with the conventional RD treatment assignment rule, where treatment is granted if the running variable exceeds the cutoff, the SABER 11 position score is multiplied by $-1$. In 2008, the eligibility cutoff was set at $-850$, while in 2009, the cutoff was raised to $-571$. Even though the eligibility cutoff changed, the amount of ACCES credit was the same across the years. This policy change establishes a non-cumulative multi-cutoff RD setup, and the extrapolation goal is to identify the treatment effect for units subject to the lower cutoff $-850$, but with running variable values lying between $-850$ and $-571$, for instance, $-650$. The key identifying assumption for extrapolation in \cite{cattaneo_keele_titiunik_2021} is the difference between the untreated potential outcome functions, defined as bias, for the two subpopulations subject to the two cutoffs is constant within the extrapolation interval (Assumption \ref{ass:constantbias}). Under this assumption, \cite{cattaneo_keele_titiunik_2021} obtain a point estimate of $0.191$ for the treatment effect for units with running variable value $-650$. Furthermore, \cite{cattaneo_keele_titiunik_2021} conduct two falsification tests to demonstrate the validity of the constant bias assumption in this application. The first one uses global polynomial regressions to test globally if the untreated potential outcome functions for two subpopulations are parallel below the low cutoff based on the regression model
\begin{equation*}
	Y_i = \alpha + \beta \mathds{1}\{C_i = h\} + \mathbf{r}_p(X_i)'\boldsymbol\gamma + \mathds{1}\{C_i = h\}\mathbf{r}_p(X_i)'\boldsymbol\delta + u_i, \mathds{E}[u_i \mid X_i, C_i] = 0,
\end{equation*}
where $\mathbf{r}_p(X_i)$ is a vector of polynomials of $X_i$ up to the $p-$th order \citep{cattaneo_keele_titiunik_2021}. The null hypothesis that corresponds to the validity of parallel trends is $\boldsymbol\delta = \mathbf{0}$. The second test uses nonparametric local polynomial methods to test the equality of the derivatives of the untreated potential outcome functions. 

\begin{figure}
	\centering
	\begin{subfigure}{.5\linewidth}
		\centering
		\includegraphics[width=8cm]{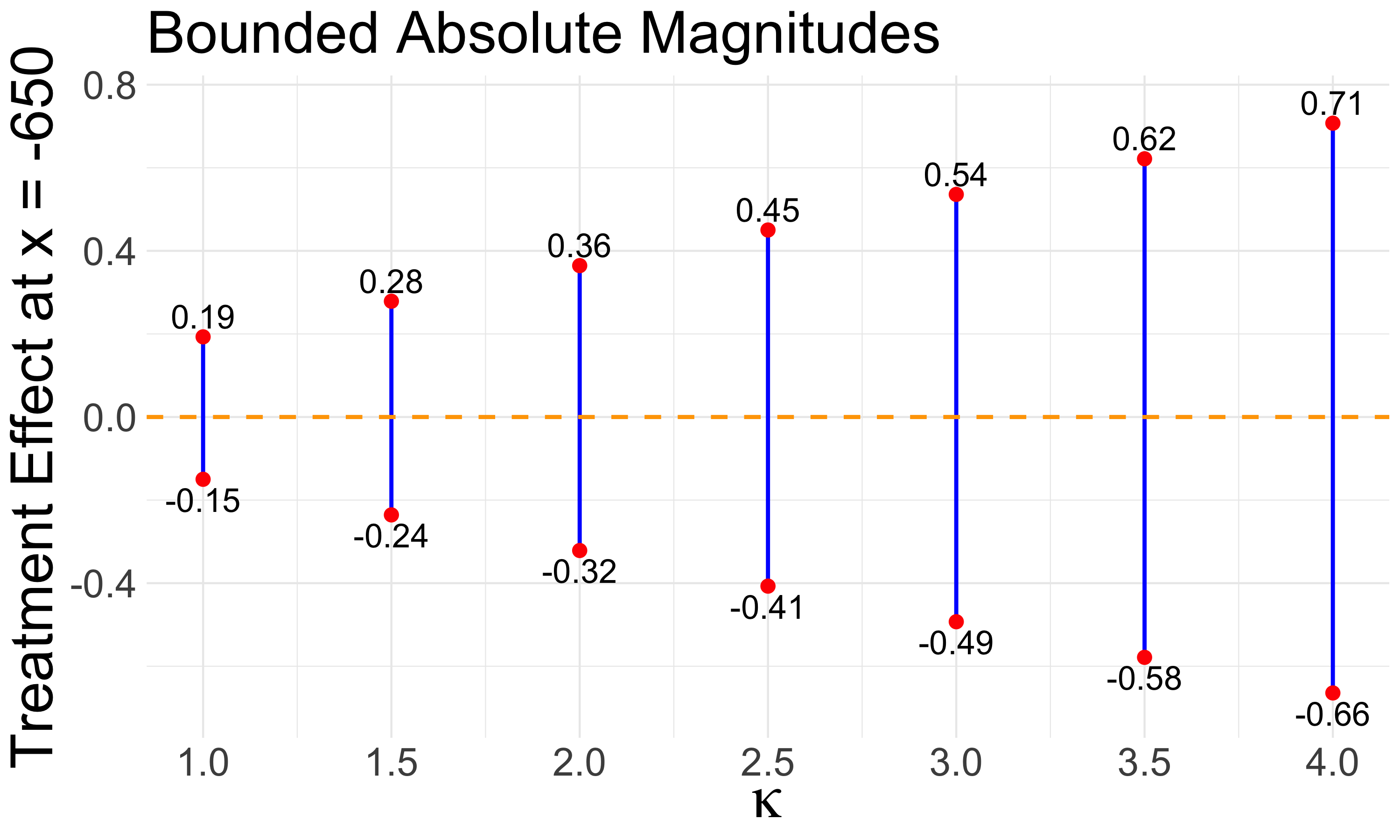}
		\caption{}
		\label{fig:jasa_am}
	\end{subfigure}%
	\begin{subfigure}{.5\linewidth}
		\centering
		\includegraphics[width=8cm]{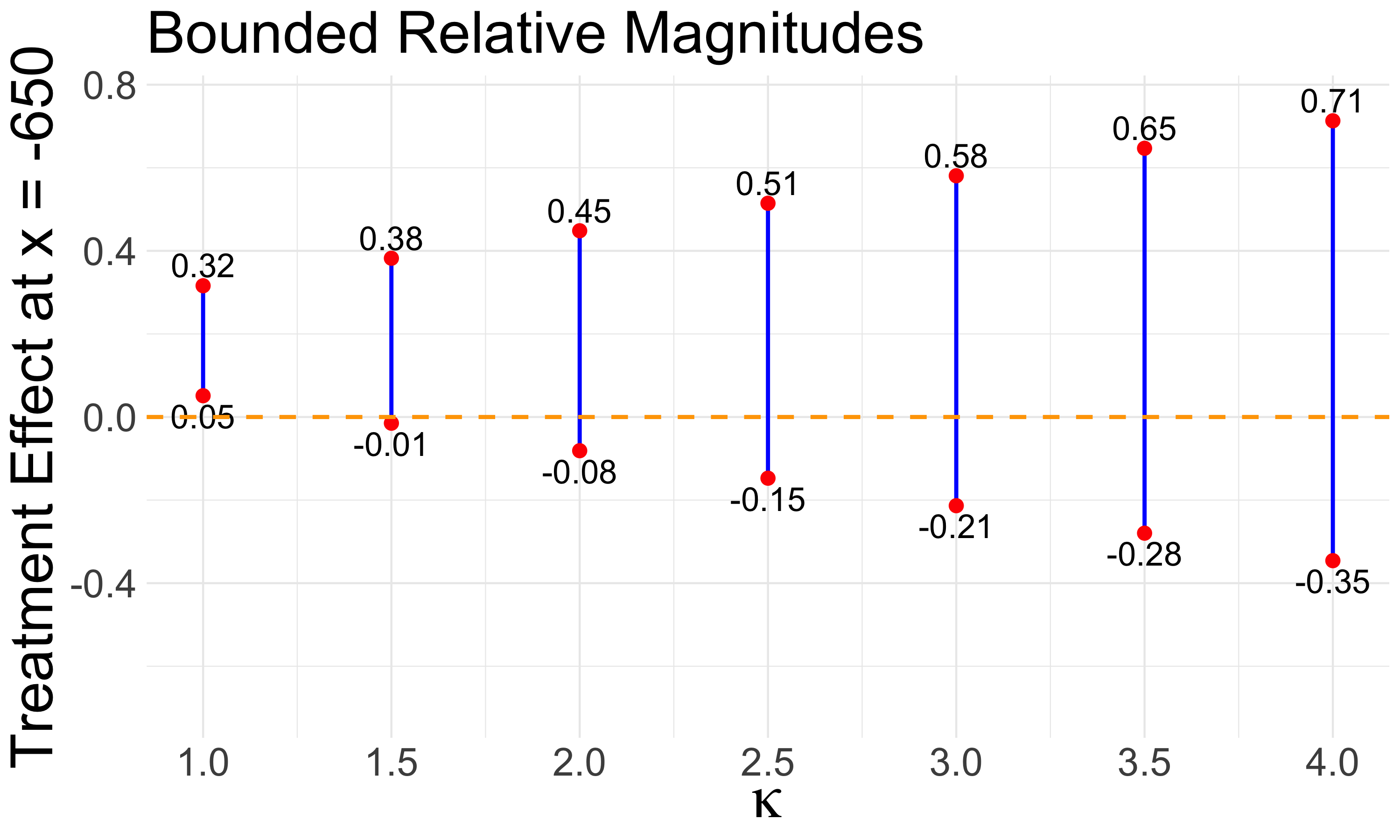}
		\caption{}
		\label{fig:jasa_rm}
	\end{subfigure}
	\begin{subfigure}{\linewidth}
		\centering
		\includegraphics[width=8cm]{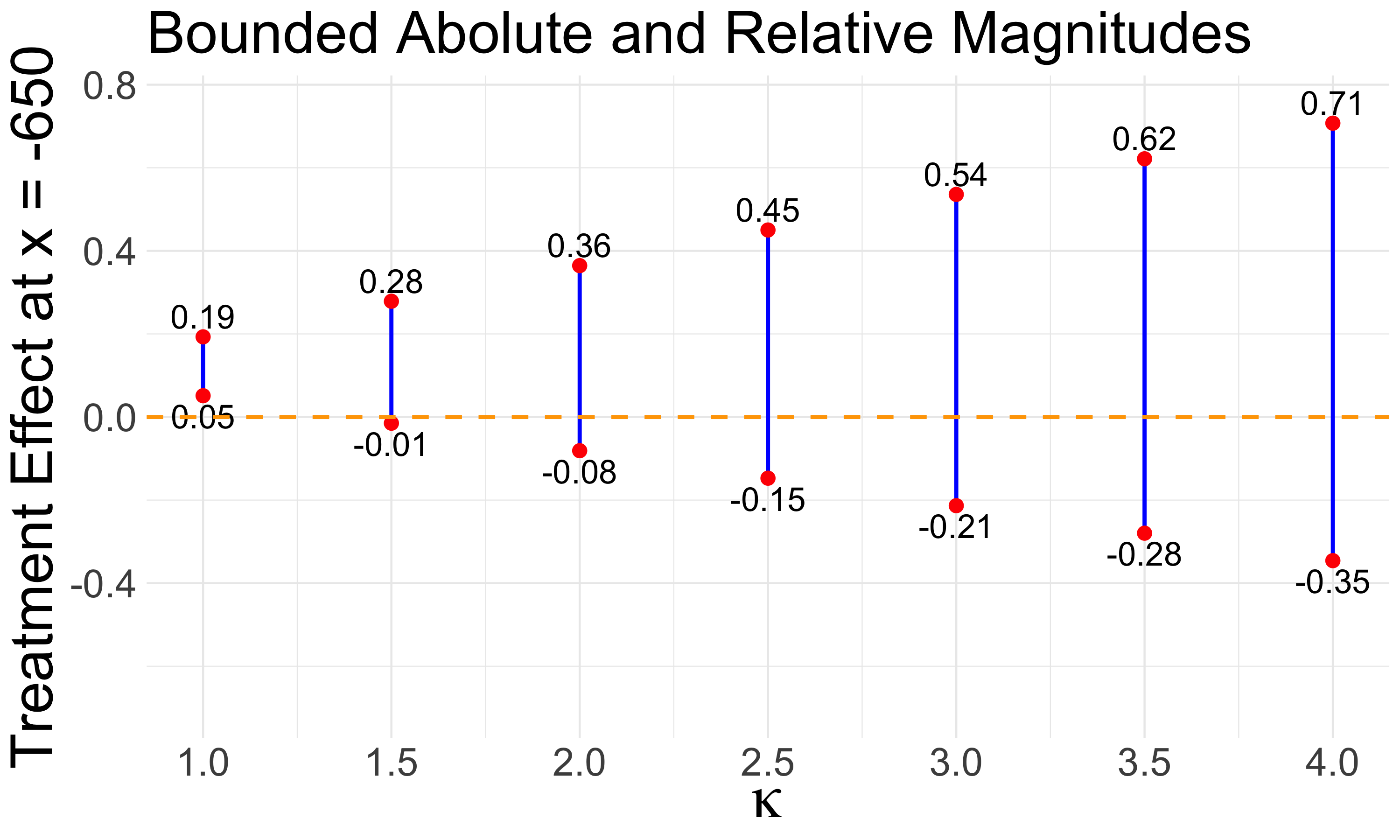}
		\caption{}
		\label{fig:jasa_amrm}
	\end{subfigure}
	\caption{Identified Sets of Treatment Effect at $x = -650$}
	\label{fig:TEjasa}
\end{figure}

Under the bounded absolute magnitude assumption, all the estimated identified sets in Figure \ref{fig:jasa_am} cover zero, which indicates no evidence to rule out zero treatment effect. Since the bias in this application is likely to be constant, restricting the magnitude of the unobserved bias to be bounded by the absolute magnitude of the maximum observed bias is not binding and does not have enough identifying power. As a sanity check, it is worth noting that in this setup, the case of constant bias corresponds to the upper bound when $\kappa=1$, which is $0.19$, and it aligns with the point estimate in \cite{cattaneo_keele_titiunik_2021}. Figure \ref{fig:jasa_rm} shows the identified sets under the bounded relative magnitude restriction. When $\kappa$ is relatively small, specifically when $\kappa = 1,$ the result shows that the partially identified set is strictly above zero, suggesting a positive treatment effect. Additionally, the point estimate $0.19$ is contained in the identification region for $\kappa = 1.$ Figure \ref{fig:jasa_amrm} shows the identified sets under both the bounded absolute and bounded relative magnitudes restriction. They have tighter bounds on the treatment effect of interest because they are the intersections of the bounds attained under each restriction. The overall result is coherent because if the true bias is constant, it requires tight restrictions on the rate of change of bias to be able to rule out zero treatment effect.

\subsection{Simulated data}

\begin{figure}
	\centering
	\begin{subfigure}{.5\textwidth}
		\centering
		\includegraphics[width=7cm]{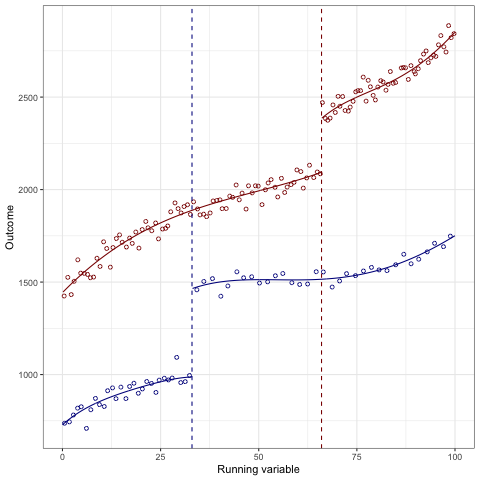}
		\caption{Multi-Cutoff RD Plot}
		\label{fig:simrdmc}
	\end{subfigure}%
	\begin{subfigure}{.5\textwidth}
		\centering
		\includegraphics[width=8.5cm]{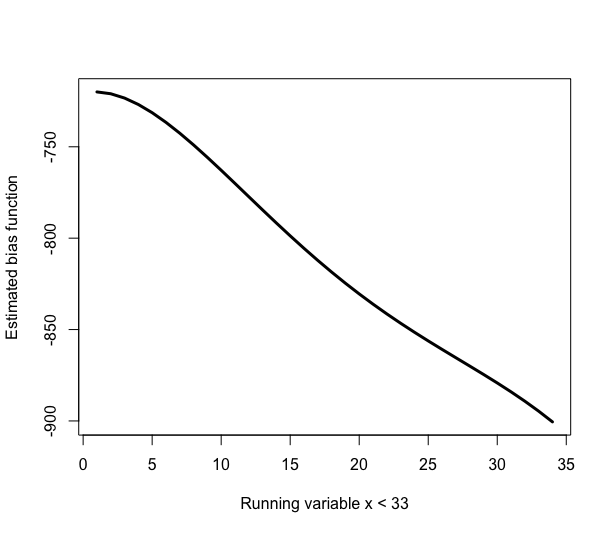}
		\caption{Estimated Bias Function}
		\label{fig:sim_biashat}
	\end{subfigure}
	\caption{Simulated Multi-Cutoff RD Design}
	\label{fig:simidset}
\end{figure}

In the next example, I extrapolate away from the cutoff in a simulated multi-cutoff design to demonstrate how the identifying power of the restrictions manifests in cases where bias is not constant. The data generating process is defined as follows.  Firstly define auxiliary random variables $\dot{X}_i$ and the cutoffs $\dot{C}_i$ as 
\begin{equation*}
	\dot{X}_i \sim Unif[0,1], \;\;\;\;\;\;\;\;\;\;\; \dot{C}_i \in \{0.33, 0.66\},
\end{equation*}
and assign half of the draws of $\dot{X}_i$ to $\dot{C}_i = 0.33$, and the rest to $\dot{C}_i = 0.66$. If $\dot{C}_i = 0.33$, generate the outcome $Y_{\ell i}$ as 
\begin{equation*}
		Y_{\ell i} = 100 \times \Big(10 + 18 (\dot{X}_i-0.5)^3 + 0.6(\dot{X}_i-0.5) + 5\times \mathds{1}\{\dot{X}_i \ge \dot{C}_i\} + \varepsilon_{\ell i} \Big), \varepsilon_{\ell i} \sim \mathcal{N}(0,1).
\end{equation*}
If $\dot{C}_i = 0.66$, generate the outcome $Y_{h i}$ as 
\begin{equation*}
	Y_{h i} = 100 \times \Big(20 + 20 (\dot{X}_i-0.5)^3 + 0.6(\dot{X}_i-0.5) + 3\times \mathds{1}\{\dot{X}_i \ge \dot{C}_i\} + \varepsilon_{h i} \Big), \varepsilon_{h i} \sim \mathcal{N}(0,1).
\end{equation*}
Lastly, scale up $\dot{X}_i$ and the cutoffs $\dot{C}_i$ by 100 as the running variable and the cutoffs, 
\begin{equation*}
	X_i =  \dot{X}_i \times 100 \;\;\;\;\;\;\; C_i = \dot{C}_i \times 100.
\end{equation*}
Figure \ref{fig:simrdmc} is a plot of the simulated data with 2000 draws from the DGP described above. The support of the running variable is $[0,100]$, and the two cutoffs are $\{33, 66\}$. The goal is to extrapolate the treatment effect for units subject to the cutoff $33$ but with a running variable value of $50$. 

Figure \ref{fig:sim_biashat} plots the estimated bias below the lower cutoff, which appears to follow a linear trend. Since the estimated bias is negative, a downward linear trend suggests that the difference between the untreated potential outcomes for the two subpopulations increases as $x$ increases. I run the two falsification tests of parallel trends discussed in \cite{cattaneo_keele_titiunik_2021}. Table \ref{tb:globalpoly} shows the results by running a global polynomial regression with maximal polynomial order of 2, which does not reject the null hypothesis of parallel trends because the coefficients of the interaction terms of $\mathds{1}\{C = h\}$ and polynomial orders of $X$ are not statistically significant. Table \ref{tb:localpoly} reports the results employing local polynomial methods. The difference between the derivatives of the untreated potential outcome functions for the two subpopulations is not statistically different from zero, which also fails to reject the parallel trend assumption. However, the true bias under this DGP is 
\begin{equation*}
	B(x) = 100 \times \Bigg(-10 - 2\Big(\frac{x}{100}-0.5\Big)^3\Bigg), \text{  for  }x \le 33,
\end{equation*}
which is clearly not constant over $x \le 33.$

\begin{table}[h] 
	\begin{center}
		\caption{Parallel trend test using global polynomial approach}
		\begin{adjustbox}{width=0.6\textwidth}
		\scriptsize
		\begin{tabular}{lcc}
			\hline
			& Estimate & Std Error\\
			\hline
			(Intercept)          & $739.15^{***}$ & $(15.31)$      \\
			$X$                   & $13.45^{***}$  & $(2.19)$       \\
			$X^2$                & $-0.18^{**}$   & $(0.07)$       \\
			$\mathds{1}\{C = h\}$          & $714.64^{***}$ & $(23.15)$      \\
			$\mathds{1}\{C = h\} \times X$        & $5.17$   & $(3.29)$       \\
			$\mathds{1}\{C = h\} \times X^2$ & $0.03$   & $(0.10)$       \\
			\hline
			$R^2$                & $0.94$    &     \\
			Adj. $R^2$           & $0.94$   &      \\
			Num. obs.            & $635$       &   \\
			\hline
			\multicolumn{2}{l}{\scriptsize{$^{***}p<0.001$; $^{**}p<0.01$; $^{*}p<0.05$}}
		\end{tabular}
		\label{tb:globalpoly}
	\end{adjustbox}
	\end{center}
\end{table}

\begin{table}[h] 	
	\begin{center}
	\caption{Parallel trend test using local polynomial approach}
	\begin{adjustbox}{width=0.618\textwidth}
	\begin{tabular}{lcccc}
		\hline
		& Estimate & Bw & p-value & 95$\%$CI \\ 
		\hline
		$\mu^{(1)}_{\ell}(\ell)$ & -0.28 & 14.25 & 0.88 & $[-60.2, 51.57]$ \\ 
		$\mu^{(1)}_{h}(h)$ & 7.81 & 19.45 & 0.00 & $[3.78, 8.65]$ \\ 
		Difference & -8.09 &  & 0.71 & $[-66.54, 45.41]$ \\ 
		\hline
	\end{tabular}
	\label{tb:localpoly}
	\end{adjustbox}
	\end{center}
\end{table}

\begin{figure}[h]
	\centering
	\begin{subfigure}{.5\textwidth}
		\centering
		\includegraphics[width=8cm]{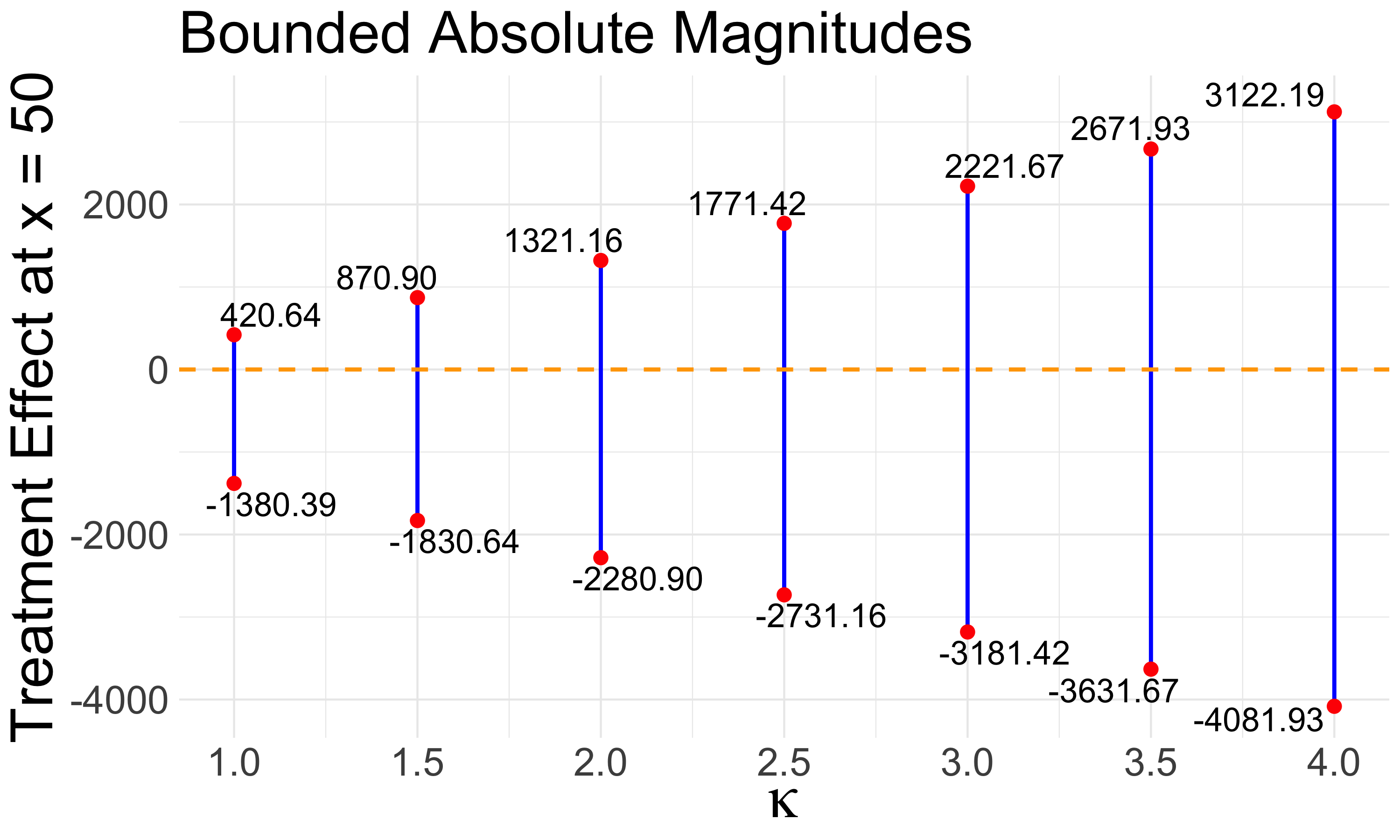}
		\caption{}
		\label{fig:sim_am}
	\end{subfigure}%
	\begin{subfigure}{.5\textwidth}
		\centering
		\includegraphics[width=8cm]{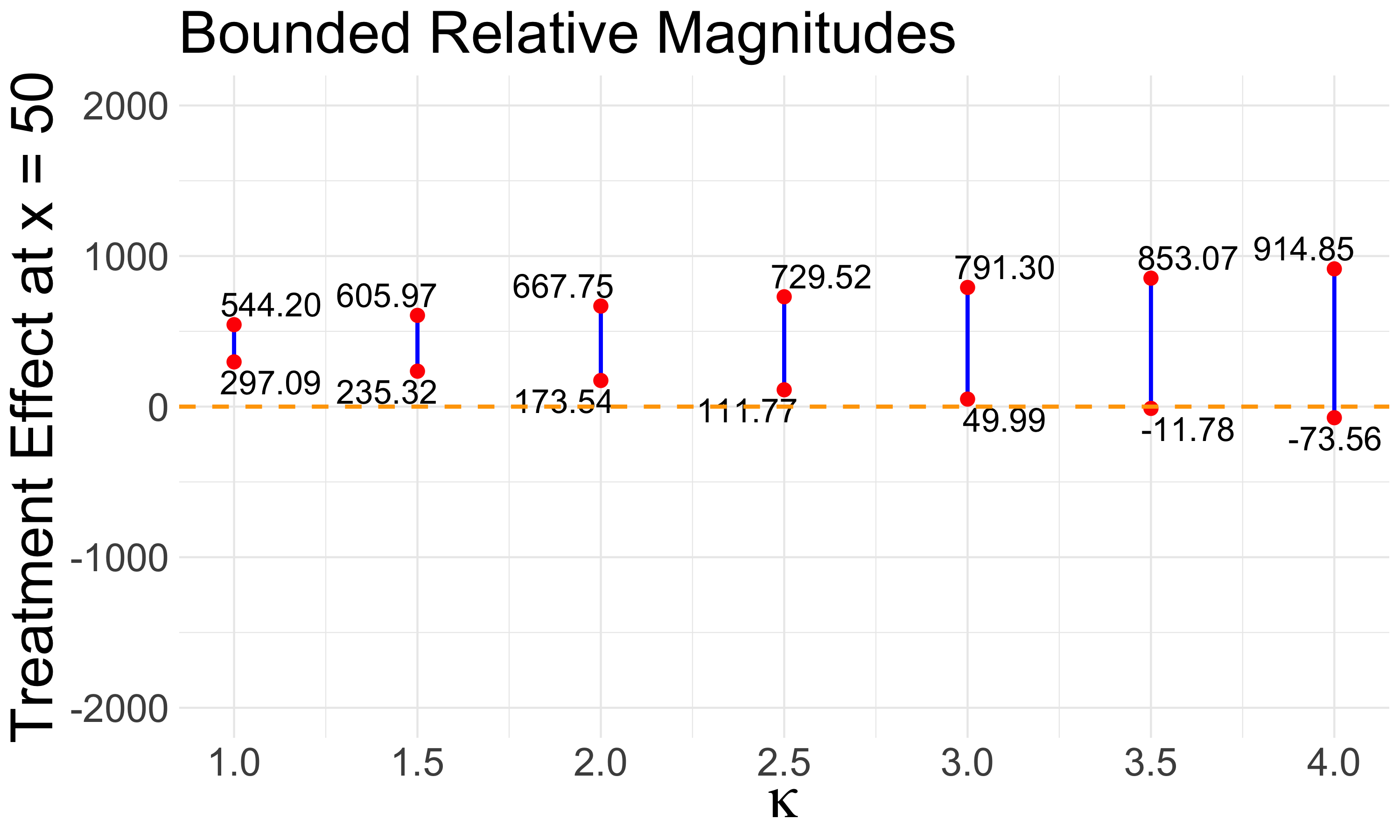}
		\caption{}
		\label{fig:sim_rm}
	\end{subfigure}
	\begin{subfigure}{\linewidth}
		\centering
		\includegraphics[width=8cm]{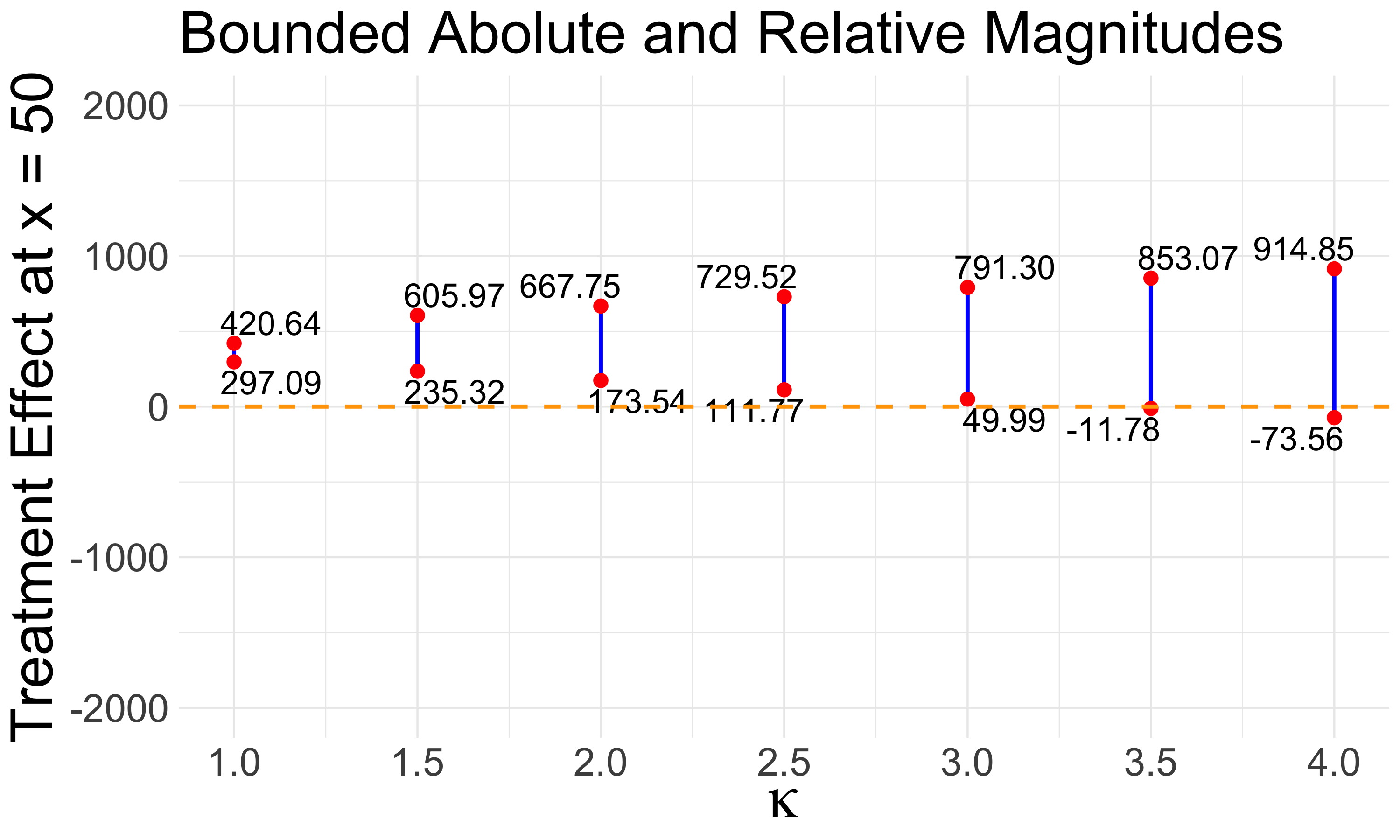}
		\caption{}
		\label{fig:sim_amrm}
	\end{subfigure}
	\caption{Identified Sets of Treatment Effect at $x = 50$}
	\label{fig:TEsim}
\end{figure}

In this case, the data fails to detect the non-constant bias below the lower cutoff. Hence, the treatment effect obtained using the approach in \cite{cattaneo_keele_titiunik_2021} is simply not valid. However, applying the method in this paper will give valid bounds on the treatment effect of interest. In this case, it is reasonable to assume bounded variation in the rate of change of the bias function for $x \in (\ell, h)$. As shown in Figure \ref{fig:sim_rm}, most of the identified sets are strictly above zero, indicating a positive treatment effect at $x = 50.$ The bounded relative magnitudes restriction imposes that the slope of the unobserved bias is not drastically different from the approximately constant slope observed in the data. Even with the assumption that the slope could be three times the magnitude of the observed slope, the resulting identified set remains strictly positive. Therefore, in this case, the researcher can conclude with greater confidence that the treatment effect at $x=50$ is indeed positive. In contrast, almost all of the identified sets under the bounded absolute magnitude restriction in Figure \ref{fig:sim_am} cover zero. Intuitively, restricting the absolute magnitude of the bias function does not capture the linear trend in bias across the running variable, and has limited identifying power, which is consistent with the observation that the bounds under the bounded absolute magnitudes assumption are much wider. Figure \ref{fig:sim_amrm} shows the identified sets by taking intersections of the bounds obtained under each restriction. Since the bounds under the bounded absolute magnitudes restriction is so wide, the intersection bounds almost perfectly overlap with the bounds under the bounded relative magnitudes restriction.

\section{Conclusion and Future Directions} \label{sec:conclusion}

This paper studies the identification problem of extrapolating the treatment effects away from the cutoff in regression discontinuity designs, where the primary challenge for identification lies in quantifying the unobserved counterfactual outcome. Instead of relying on strong assumptions that point-identify the treatment effect but may result in identification failure in the case of slight violations, this paper proposes a non-exhaustive list of bounded variation assumptions on the evolution of the counterfactual outcome function along the running variable. This approach achieves partial identification of the treatment effect of interest and encompasses both single and multiple cutoff designs. Through simulations, the paper demonstrates that by applying reasonable restrictions based on the specific empirical setting, researchers can derive credible and informative bounds of the treatment effect away from the cutoff.

There are two primary directions for future research stemming from this paper. One ongoing effort involves establishing the statistical properties of the identification region and conducting inference, which will be built on previous work by \cite{imbens_manski_2004} and \cite{stoye_2009}.  Another direction pertains to exploring the policy implications that arise from partially identified treatment effects away from the cutoff. This could involve investigating whether the status-quo policy is optimal and whether policymakers should consider adjusting the cutoff in order to maximize overall social welfare.

\bibliographystyle{apalike}
\bibliography{ref}


%

\end{document}